\documentclass[12pt,a4paper]{article}
\usepackage{amsfonts}
\usepackage{amssymb}
\usepackage{latexsym}
\begin{document}

\begin{center}
{\Large \bf Maximally symmetric spaces in
Brans-Dicke theory\\ and the cosmological constant} \\

\vspace{4mm}

Mikhail N. Smolyakov\\ \vspace{0.5cm} Skobeltsyn Institute of
Nuclear Physics, Moscow State University
\\ 119991
Moscow, Russia\\
\end{center}

\begin{abstract}
In this note we discuss a toy model in which the value of the
effective cosmological constant is associated with a symmetry
properties of the vacuum background solution and defined by the
model parameter associated with a scalar field non-minimally
coupled to gravity. Although the parameters in the model seem to
have unrealistic values to account for the present value of the
cosmological constant, the model demonstrates a possible
alternative way it can arise.
\end{abstract}

During the last years the problem of cosmological constant, its
origin and small value attracts much attention. It is possible
that the vacuum energy is indeed a constant, and there are a lot
of attempts to explain the existence of such constant vacuum
energy -- see, for example, reviews \cite{Review1,Reviews} and
references therein. One of the most interesting questions is that
about its extremely small value. Indeed, naive consideration show
that the vacuum energy value should be many orders of magnitude
larger. In this note we discuss a toy model, which does not solve
this problem, but proposes a possible way to change the origin of
the cosmological constant.

First, let us consider the action
\begin{eqnarray}\label{gravact0}
S=\int d^{4}x\sqrt{-g}\left[\phi
R-\omega\frac{\partial^{\mu}\phi\partial_{\mu}\phi}{\phi}-V(\phi)\right],
\end{eqnarray}
where $R$ is the four-dimensional curvature, $\phi$ is the
Brans-Dicke field, $\omega$ is the dimensionless Brans-Dicke
parameter. The equation of motion for the Brans-Dicke field
$\phi$, coming from this action, has the form
\begin{equation}\label{BDEQ0}
\left(2\omega+3\right)\nabla^{\mu}\nabla_{\mu}\phi+2
V(\phi)-\phi\frac{d V(\phi)}{d\phi}=0,
\end{equation}
where $\nabla_{\mu}$ is the covariant derivative with respect to
the metric. For the case $V(\phi)=0$ we get the well known
equation of the original Brans-Dicke theory \cite{Misner:1974qy}
\begin{equation}\label{BDEQ1}
\left(2\omega+3\right)\nabla^{\mu}\nabla_{\mu}\phi=0.
\end{equation}
In this case the static background solution of Einstein and scalar
field equations has the form
\begin{eqnarray}
g_{\mu\nu}=\eta_{\mu\nu},\\
\phi=const,
\end{eqnarray}
where $\eta_{\mu\nu}$ is the Minkowski metric.

Now let us consider another possibility. Namely, let us take
\begin{equation}\label{V}
V(\phi)=\gamma\phi^{2},
\end{equation}
where $\gamma$ is a dimensionless constant (cosmology of
Brans-Dicke theory with such form of the potential was discussed
in \cite{Santos:1996jc}). The scalar field equation for the choice
(\ref{V}) again takes a simple form (\ref{BDEQ1}). It is evident
that there exists a static solution of equation (\ref{BDEQ1})
$\phi=const$. As for the metric, corresponding background
solutions take the form of $dS_{4}$ for $\gamma>0$ and $AdS_{4}$
for $\gamma<0$, which can be easily seen from the Einstein
equations \cite{Misner:1974qy}:
\begin{eqnarray}\label{Eeq}
\phi\left(R_{\mu\nu}-\frac{1}{2}g_{\mu\nu}R\right)=-\frac{1}{2}g_{\mu\nu}\gamma\phi^{2}+
\\ \nonumber
+\frac{\omega}{\phi}\left(\nabla_{\mu}\phi\nabla_{\nu}\phi-\frac{1}{2}g_{\mu\nu}
\nabla^{\rho}\phi\nabla_{\rho}\phi\right)+\nabla_{\mu}\nabla_{\nu}\phi-g_{\mu\nu}
\nabla^{\rho}\nabla_{\rho}\phi.
\end{eqnarray}
Thus, regardless of the sign of $\gamma$, there exist solutions
which describe maximally symmetric spaces (which are Minkowski for
$\gamma=0$, $dS_{4}$ or $AdS_{4}$ for $\gamma\ne 0$). This result
is well known. Indeed, passing from the Jordan frame to the
Einstein frame with the help of conformal transformations results
in reducing the scalar field potential to the constant potential
$V^{*}(\phi)\sim\gamma$. An important point is that though
solutions in different frames are equivalent, the corresponding
theories in different frames are not equivalent, because if in the
Brans-Dicke theory we had the minimal coupling of gravity to any
additional matter, then after the conformal transformation the
scalar field would enter the term describing the interaction with
matter (the so-called conformal ambiguity \cite{Overduin:1998pn}).
Thus, if we suppose that the physical metric is that of the Jordan
frame, it is more convenient to use original action of the form
(\ref{gravact0}).

The situation radically changes if $V\ne \gamma\phi^{2}$, for
example, if $V= \gamma\phi^{2}+\Lambda$, where $\Lambda$ is a
constant. One can check that in this case $\phi\ne const$ and
solution for the metric does not correspond to a maximally
symmetric space. Thus, a maximally symmetric spaces can be
realized only if $\Lambda=0$ (we note, that in the absence of the
Brans-Dicke scalar field, i.e. in the standard case, the maximally
symmetric spaces exist for any value of $\Lambda$). But it is not
evident why it should be so. Nevertheless, one can recall that
there exists the well-known mechanism based on introducing the
3-form gauge field into the theory
\cite{Aurilia:1980xy,Aurilia:1980xj,Henneaux:1984ji,Hawking:1984hk},
which turns the cosmological constant into an integration
constant.

To this end let us consider the action of the form
\begin{eqnarray}\label{gravact}
S=\int d^{4}x\sqrt{-g}\left[\phi
R-\omega\frac{\partial^{\mu}\phi\partial_{\mu}\phi}{\phi}-\gamma\phi^{2}-\bar\Lambda-
\frac{1}{48}F_{\mu\nu\rho\sigma}F^{\mu\nu\rho\sigma}+L_{matter}\right],
\end{eqnarray}
where $\bar\Lambda>0$ is the "bare" energy density of the vacuum
and supposed to include, for example, the contribution of quantum
fluctuations, thus its value can be large, $L_{matter}$ stands for
the Lagrangian of other matter (we would like to mention again
that the metric in this frame is supposed to be the physical
metric). The 3-form gauge field arises in supergravity theories
\cite{Aurilia:1980xy}, but in the toy model under consideration it
is introduced without reference to any particular theory.

The solution of equations of motion for the field strength of the
3-form gauge field is
\begin{equation}\label{3formeq}
F^{\mu\nu\rho\sigma}=\frac{c\epsilon^{\mu\nu\rho\sigma}}{\sqrt{-g}},
\end{equation}
where $c$ is a constant, and the contribution of this 3-form field
to the action reduces to
\begin{equation}
-\frac{1}{48}F_{\mu\nu\rho\sigma}F^{\mu\nu\rho\sigma}\to
+\frac{c^{2}}{2}.
\end{equation}
The constant $c$ in (\ref{3formeq}) is not fixed by the equations
of motion for the 3-form field. Solution (\ref{3formeq}) is valid
for any metric, and once the constant $c$ is fixed, it remains
unchanged. The 3-form field appears to be non-dynamical.

The equation of motion for the Brans-Dicke field $\phi$
(\ref{BDEQ0}) now takes the form
\begin{equation}\label{BDEQ3}
\left(2\omega+3\right)\nabla^{\mu}\nabla_{\mu}\phi+2\bar\Lambda-c^{2}=8\pi
T.
\end{equation}
In principle the constant $c$ can be arbitrary, leading to an
arbitrary value of the effective energy density
$\bar\Lambda-c^{2}/2+\gamma\phi^{2}$. There are several methods to
fix its value in the theory without non-minimally coupled scalar
field, see, for example, \cite{Review1}, resulting in the
vanishing of the effective cosmological constant. Here we would
like to propose another possible method. Indeed, since the
cosmological constant is treated as a vacuum energy density, let
us consider a vacuum solutions for system (\ref{gravact}) (i.e. in
the case $T_{\mu\nu}=0$). There are different solutions depending
on the value of $c$, mainly with time-dependent $\phi$ and,
consequently, time-dependent metric (we are interested in
isotropic solutions). But there is also a class of solutions,
which possesses an additional property -- a symmetry. Thus, we
apply the symmetry criterion for choosing the value of constant
$c$: we suppose that the (global) vacuum background solution is
maximally symmetric. We note that we do not propose any particular
mechanism leading to such choice of constant $c$. Moreover it is
not easy to justify this ad hoc hypothesis, but in this connection
one can recall the ideas that the Universe could have global
topology and structure, which were discussed, for example, in
\cite{Luminet,Bogoslovsky}. Nevertheless, our hypothesis is
nothing but an assumption, which can not be consistently justified
within the framework of classical theory, and one can only suppose
that some more general theory could deal with symmetry properties
of a space-time.

The proposed assumption uniquely defines the value of $c$ (since
the scalar field $\phi$ should be constant for the maximally
symmetric solutions):
\begin{equation}\label{DE}
c^{2}=2\bar\Lambda.
\end{equation}
Thus, the 3-form field totally compensates the contribution of
$\bar\Lambda$ to the energy density of the vacuum. It is evident
that the background value of the scalar field should be identified
with the Planck mass, $\phi=M_{Pl}^{2}=const$. If $\gamma>0$, we
would get $dS_{4}$ background solution, i.e. the case of positive
cosmological constant. In the limit $x^{0}\to\infty$ the ordinary
and dark matter average densities tend to zero, $\phi\to
M_{Pl}^{2}$, solutions for the metric and the scalar field tend to
this vacuum solution. Of course the existence of matter, described
by $T_{\mu\nu}$, breaks the symmetry of the vacuum background
solution.

Now the contribution to the effective cosmological constant is
defined by the term $\gamma\phi^{2}$ and $$\rho_{vac}\sim\gamma
M_{Pl}^{4}.$$ Of course, the value of $\gamma$ should be extremely
small to reproduce the present value of the cosmological constant.
Such a small value seems to be unrealistic. Moreover, the problem
of the small value of parameter $\gamma$ is not better than the
original cosmological constant problem. There are other
disadvantages of this scenario. For example, the mechanism
discussed above works only for the case $\bar\Lambda=const$. We
also do not discuss in this note the mechanism which sets the
vacuum expectation value of the Brans-Dicke field. It is possible
that there could appear a mass term $\sim M_{Pl}^{2}$ for the
field $\varphi=\phi-M_{Pl}^{2}$, in this case the backreaction of
the Brans-Dicke field on the metric fluctuations appears to be
totally negligible. But one should take into account that the
simple model presented above is nothing but a toy model, which
nevertheless can be interesting from the theoretical point of
view, because it demonstrates that there could be a connection
between a value of the effective cosmological constant and a
symmetry properties of the space-time, leading to a possibility
that the effective cosmological constant is defined by some theory
non-minimally coupled to gravity, whereas the proper energy
density of the vacuum $\bar\Lambda$ does not contribute to the
effective cosmological constant at all.

\section*{Acknowledgements}
The author is grateful to G.Yu.~Bogoslovsky and
I.P.~Volobuev for valuable discussions. The work was supported by
grant of Russian Ministry of Education and Science NS-1456.2008.2,
grant for young scientists MK-5602.2008.2 of the President of
Russian Federation, grant of the "Dynasty" Foundation and
scholarship for young scientists of Skobeltsyn Institute of
Nuclear Physics of M.V.~Lomonosov Moscow State University.

\end{document}